\documentclass[10pt,twocolumn,a4paper,fleqn]{article}

\usepackage[sectionbib,square]{natbib}

\usepackage{balance}
\newcommand{\myemail}{{\let\thefootnote\relax\footnote{$\star$
    \texttt{Lukasz.Bratek@ifj.edu.pl}}}}
\newcommand{\mytitle}[1]{\begin{quotation}{\bf\huge\noindent #1}
    \end{quotation}}
\newcommand{\myabstract}[1]{\begin{quotation}\noindent{\bf Abstract.}{
    \small #1}\end{quotation}}

\usepackage[active]{srcltx}
\usepackage[dvips]{graphicx}
\usepackage{amsmath}
\usepackage[scriptsize,bf]{caption}
\usepackage{float}
\usepackage{placeins}

\usepackage[polish,english]{babel}
\usepackage{xspace}

\newcommand{\ud}[1]{\mathrm{d}#1}

\newcommand{\br}[1]{\left(#1\right)}
\newcommand{\abs}[1]{\left|#1\right|}

\newcommand{\figref}[1]{Fig.\ref{#1}}
\renewcommand{\eqref}[1]{Eq.\ref{#1}}
\newcommand{\eqrefs}[1]{Eqs.\ref{#1}}

\newcommand{\kpc}{\,\mathrm{kpc}}
\newcommand{\pc}{\mathrm{pc}}
\newcommand{\km}{\mathrm{km}}

\newcommand{\erf}[1]{\mathrm{erf}\br{#1}}
\newcommand{\muG}{\,\mu \mathrm{G}}
\newcommand{\hi}{H{\scriptsize I}\xspace}
\newcommand{\hii}{H{\scriptsize II}\xspace}

\usepackage{color}
\newcommand{\corr}[1]{{#1}}
\newcommand{\new}[1]{{\color{blue}#1}}
\usepackage[normalem]{ulem}

\begin{document}
\twocolumn[\begin{@twocolumnfalse}

\bigskip\bigskip\bigskip

\mytitle{Magnetic flux density from the relative circular motion
of stars and partially ionized gas in the Galaxy mid-plane
vicinity}

\medskip

\begin{center}
{ \large Joanna
Ja{\l}ocha$^{1}$, {\L}ukasz Bratek$^{2,\star}$,  Jan P\c{e}kala$^{2}$, Szymon Sikora$^{3}$, Marek Kutschera$^{4}$}\\
\medskip
\begin{tabular}{@{}l}
{\small 1. Institute of Physics, Cracow University of Technology, PL-30084 Krakow, Poland}\\
{\small 2. Institute of Nuclear Physics, Polish Academy of Sciences, PL-31342 Krakow, Poland}\\
{\small 3. Astronomical Observatory, Jagellonian University, PL-30244 Krakow, Poland}\\
{\small 4. Institute of Physics, Jagellonian University, PL-30059 Krakow, Poland}\\
\end{tabular}\\
\bigskip \texttt{Preprint v1, 24 OCT 2016, IFJPAN-IV-2016-25}
\end{center}

\myabstract{Observations suggest  a slower stellar rotation relative to gas
rotation in the outer part of the Milky Way Galaxy. This
difference could be attributed to an interaction with the
interstellar magnetic field. In a simple model, fields of order
$10\muG$ are then required, consistently with the observed
values. This coincidence suggests a tool for estimating magnetic
fields in spiral galaxies. A North-South asymmetry in the rotation
of gas in the Galaxy could be of magnetic origin too.}


\bigskip
  \end{@twocolumnfalse}]
\medskip


\section{Introduction}\myemail

The mechanism of the influence of turbulent and large-scale
magnetic fields on the motion of the gaseous fraction was
described in the context of galactic dynamics  by
\citet{bib:nature}. The interstellar gas is ionized to a degree
sufficient for the magnetic field to freeze-in the gas, so that
the resulting magnetic tension can be assumed to hold the gas and
so influence its motion, {modifying} the predictions of purely
gravitational models. Because the gas density decreases with the
galactocentric distance, the magnetic effect becomes important for
larger radii.

There have been several works devoted to this problem so far,
showing that magnetic fields of a few$\muG$ (typical for the
interstellar medium in spiral galaxies) are capable of
considerably influencing the kinematics of (at least) partially
ionized gas \citep{bib:nature,bib:battaner,bib:kutschera,
bib:battaner2,bib:polema1,bib:polema2,bib:polenasze2,bib:polenasze1}.
The fields have the potential to modify the outer parts of gaseous
rotation curves, and in turn, the predictions about the
distribution of mass.

Magnetic fields can not be ignored when one attempts to fully
understand the problem of rotation of spiral galaxies.
\citet{bib:nature} brought to attention the possibility that the
observed flatness of outer disc rotation curves, or even their
rise, could be simply the result of interaction with interstellar
magnetic fields (not merely due to unseen dark matter halo). In
particular, it might turn out, that taking magnetic fields into
account would reduce the amount of non-baryonic dark matter
required by models that ignore magnetic fields.

Interestingly, using position-velocity data for a sample of remote
classical cepheids and for \hii regions in the Galaxy,
\citet{1997A&A...318..416P} found that the rotation curve
indicated by the stars in the outer disc is markedly lower than
the rotation curve of gas in the same region. They attributed
{this difference} to either non-axisymmetric components in the gas
motion, or to high uncertainties in the distances to \hii regions.
From \citep{1997A&A...318..416P} it is evident also that there is
a South-North asymmetry of the gaseous rotation curve, not
observed for the stellar rotation curve. Remarkably, {the
difference in the rotation curves} concerns the outer part of the
Galactic disc and increases with the Galactocentric distance. The
{difference} seems too high to be solely due to the neglected
velocity components. Rather, the differences may be a
manifestation of the influence of magnetic fields acting upon the
gaseous substructure, increasing the rotation of gas with respect
to that of stars. By neglecting magnetic fields, this increase
would be customarily attributed to higher amounts of dark matter
distributed in outer regions.

So far, the studies on the influence of magnetic fields on
rotation curves have been inconclusive. Some authors, e.g.
\citet{bib:battaner2,bib:polema2}, state that magnetic fields are
responsible for the flattening or rise of outer parts of rotation
curves, other authors provide sound arguments that the influence
is unlikely, or even may impede the gravitationally supported
rotation, leading to an even higher halo mass \citep{bib:magp,
bib:magp2}.  However, the effects are subject to modelling
assumptions.

The motivation behind the present work is our conjecture, that the
observed {difference} between the stellar and gaseous rotation
curves of the Milky Way galaxy can be explained by the influence
of magnetic fields. With this assumption, we estimate the
magnitude and profiles of the radial and azimuthal magnetic field
components required to account for the difference in rotation.

\section{Gaseous and stellar rotation curves}
The Milky Way rotation curve is determined based on the
measurements of various kinematical tracers moving in the Galactic
plane vicinity. To obtain the stellar and gaseous  rotation
curves, we unified position-velocity data for $357$ cepheids
\citep{1987AJ.....93.1090C, 1994A&A...285..415P,
1997A&A...318..416P, 2000A&AS..143..211B, 2015AN....336...70M},
$110$ carbon stars \citep{2007A&A...473..143D,
2013Ap.....56...68B} and $255$ \hii regions
\citep{1982ApJS...49..183B, 1984A&A...139L...5C,
1989ApJ...342..272F, 1993A&A...275...67B}. The data were
originally used as tracers of rotation in the {Galactic} plane
vicinity, therefore we do not perform any further selection. We
transformed the radial distances, radial motions and transverse
motions (and the respective errors) from the heliocentric
coordinate frame to the Galactocentric coordinate frame,  assuming
the IAU Galactic constants $R_o=8.5\kpc$ and
$V_o=220\,\km/\!\sec$.\footnote{Some of the data had to be
recalculated from other Galactic constants; e.g., $R_o=7.6\kpc$
used in \citep{2013Ap.....56...68B}} For each data point we found
the projected distance and the azimuthal velocity component in the
{Galactic} plane (that is, $R$ and $R\cdot\ud{\Phi}/\ud{t}$ in
cylindrical coordinates). We denote them as $(r_i,v_i)$ pairs.

\begin{figure}
   \centering
   \includegraphics[width=0.475\textwidth  ,trim={0cm 0cm 0cm 0cm}]{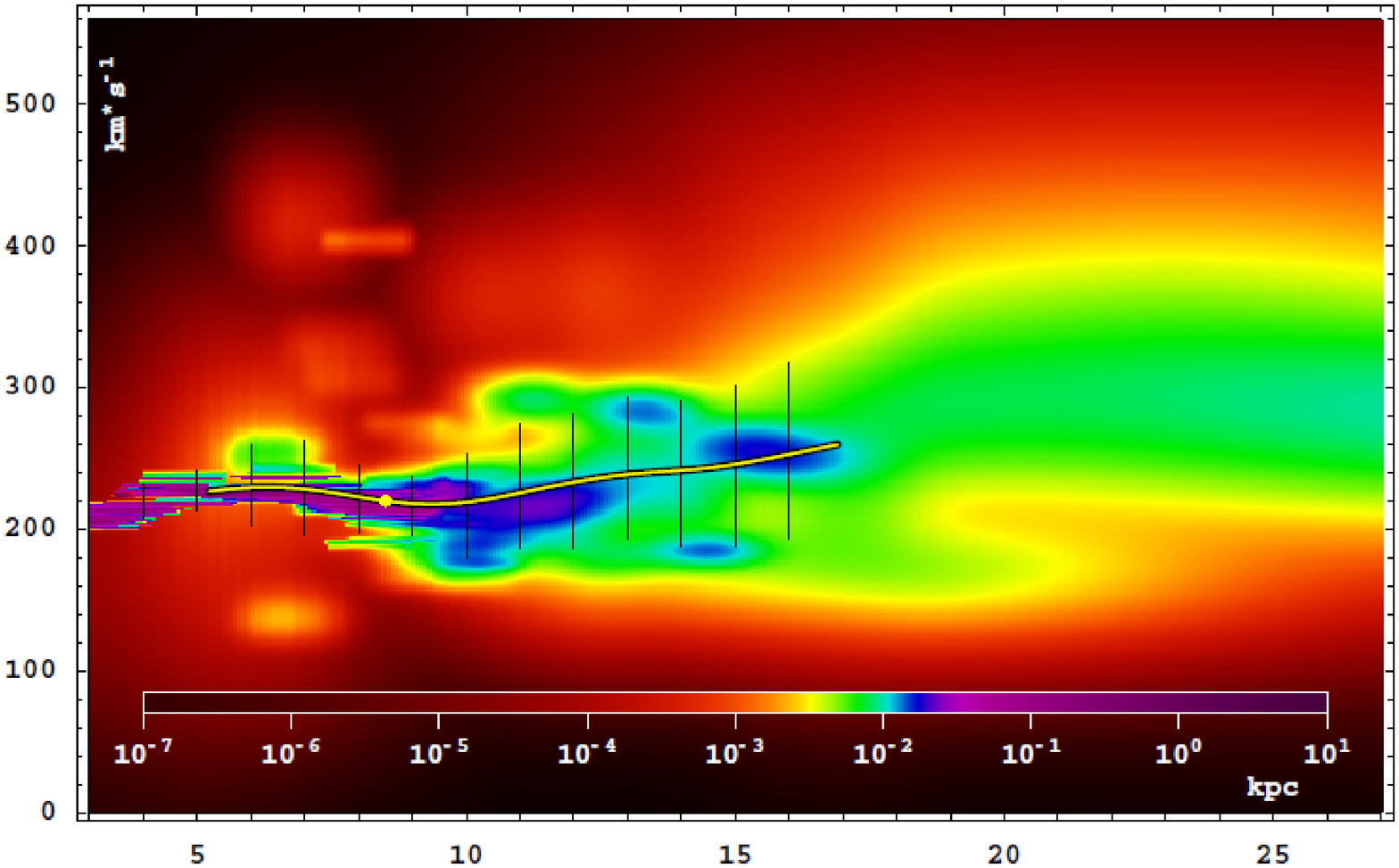}
   \includegraphics[width=0.475\textwidth,trim={0cm 0cm 0cm 0cm}]{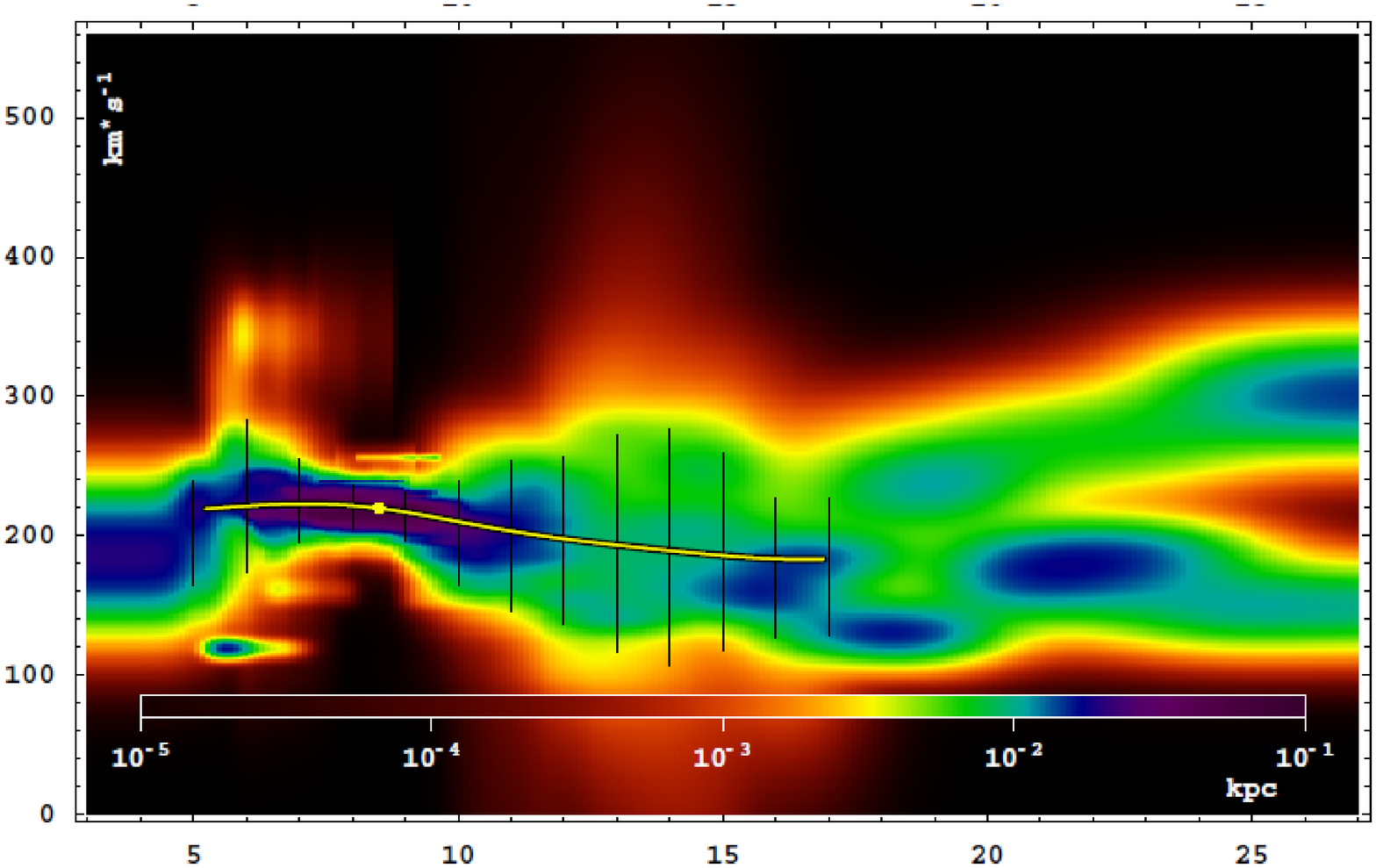}
\includegraphics[width=0.415\textwidth,trim={0cm 0cm 0cm 0cm}]{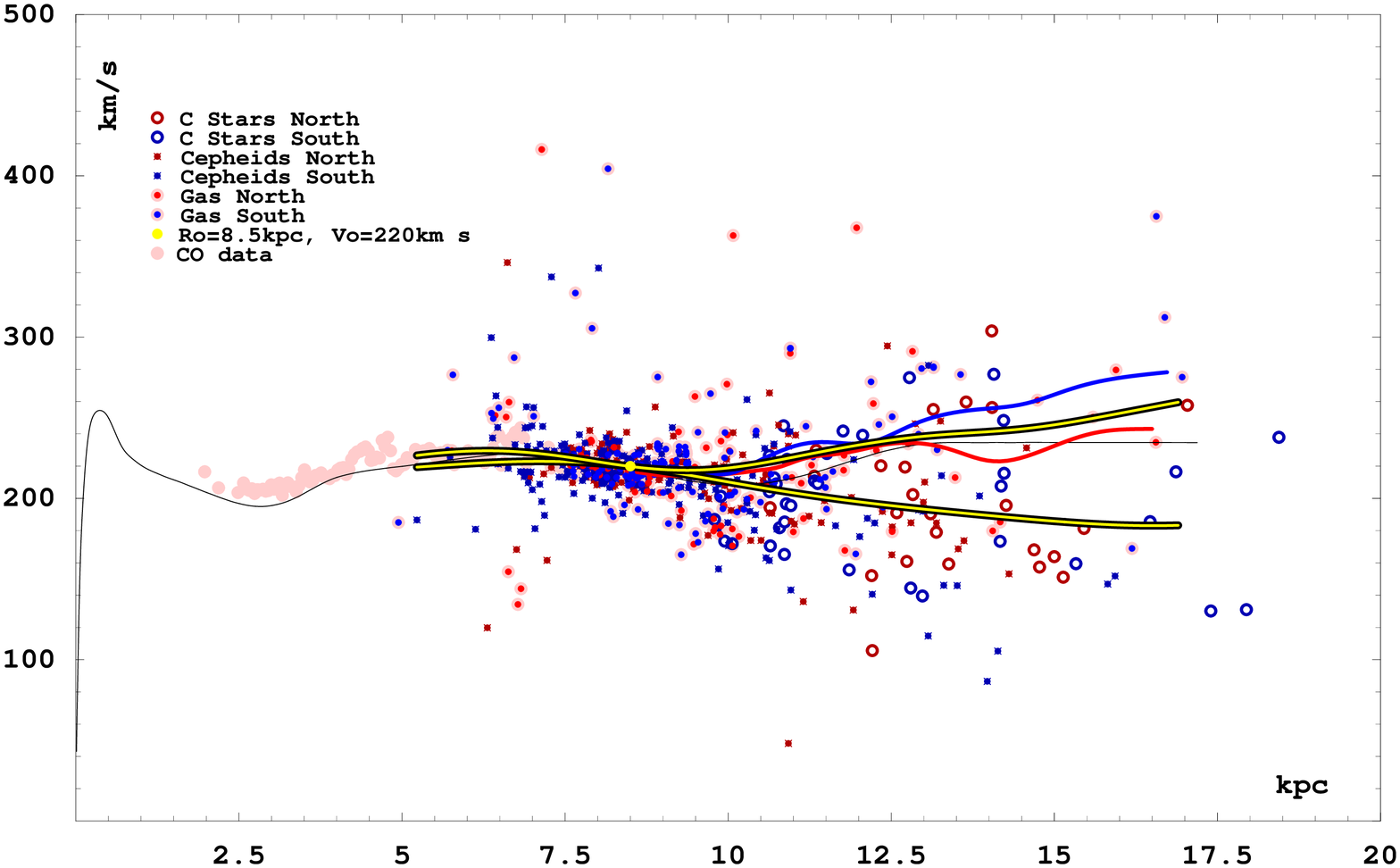}
      \caption{\label{fig:RC}  Conditional distribution function for
position-velocity measurements in Galactic coordinates (assuming
IAU Galactic constants $R_o=8.5\kpc$ and $V_o=220\,\km/\!\sec$,
indicated on the graphs with the yellow dot), for gas [{\it top
panel}] and for stars (unified data for cepheids and carbon stars)
[{\it middle panel}] respectively, which we used to obtain gaseous
and stellar rotation curves shown here with black-yellow lines
{(the vertical thin lines represent a region within a
deviation $S_V(R)$ from the mean $E_V(R)$)}.
The position-velocity points are shown explicitly [{\it bottom
panel}], together with the stellar and gaseous rotation curves
from upper panels. The {\it red solid line} and the {\it blue
solid line} show the rotation curves for the gas above and below
the {Galactic} plane, respectively. To verify our method of
obtaining the rotation curves, we show  with the {\it black thin
line} an independent gaseous rotation curve of Galaxy,
\corr{reproduced from} a polynomial fit to gas position-velocity
data published by \citet{1985ApJ...295..422C} (assuming the same
IAU Galactic constants). }
  \end{figure}

To obtain a rotation curve from a set
$\mathcal{S}=\{(r_i,v_i)\}_{1\leq i\leq n}$, we start with a
distribution function
$\tilde{p}(r,v)=\tilde{\mathcal{N}}_S^{-1}\sum_{i=1}^{n}\exp\br{-\frac{(r-r_i)^2}{2\br{\Delta
r_i}^2}-\frac{(v-v_i)^2}{2 \br{\Delta v_i}^2}}$,
$\tilde{\mathcal{N}}_\mathcal{S}$ being an $\mathcal{S}$-dependent
normalization constant, and $\Delta r_i$ and $\Delta v_i$  the
measurement uncertainties. To simplify calculations without
noticeably changing the results, the integration regions were
formally extended to $\pm\infty$ in both $R$ and $V$ variables,
because the summands in $\tilde{p}$ are rapidly decreasing outside
a convex region encompassing entire $\mathcal{S}$. Next, we form a
related smoothed-out distribution function $p$ by integrating
$\tilde{p}$ within intervals $(R-w/2,R+w/2)$:
$p(R,V)=\mathcal{N}_S^{-1}\sum_{i=1}^{n}\exp\br{-\frac{(V-v_i)^2}{2
\br{\Delta v_i}^2}}f_{w,i}(R)$ where $f_{w,i}(R)=\frac{1}{2w}
\br{\erf{\frac{R-r_i+w/2}{\sqrt{2}\Delta r_i}}
     -\erf{\frac{R-r_i-w/2}{\sqrt{2}\Delta r_i}}}
$ are integrable to $1$ on the interval
$R\in\br{-\infty,+\infty}$. Then, corresponding to $p(R,V)$, the
conditional expectation value of $V$ denoted by $E_V(R)$ and its
\corr{dispersion measure} $S_V(R)$ read {(the summation is
taken over all tracers)}:
\begin{eqnarray*} E_V(R)&=&\frac{\sum_{i=1}^{n}v_i f_{w,i}(R)}{\sum_{i=1}^{n}f_{w,i}(R)},\\
S_V(R)&=&\sqrt{\frac{\sum_{i=1}^{n}\br{\br{E_V(R)-v_i}^2+\br{\Delta
v_i}^2}f_{w,i}(R)}{\sum_{i=1}^{n}f_{w,i}(R)}}\corr{.}\end{eqnarray*}
{Note, that} the $S_V(R)$ involves both the uncertainties $\Delta v_i$ and
$\Delta r_i$. We chose the window width $w\approx1.5\kpc$ for so
defined moving averages $E_V$ and $S_V$. The moving average lines
were then smoothed-out using a smoothing spline interpolation
method. So obtained gaseous and stellar rotation curves are
plotted in \figref{fig:RC}, together with their respective
conditional density distributions.

For the purpose of this paper, the stellar and gaseous rotation
curves were made to coincide at the point $(R_o,V_o)$, even though
fractions of galactic material in the neighborhood of the Local
Standard of Rest perform a small relative motion. But, if we are
to compute the magnetic field necessary to obtain a given
difference between two rotation curves, first we have to exclude
any contribution to the actual difference, which is known to be of
non-magnetic origin. The latter will certainly be dominating,
because the density of gas in the vicinity of the circular orbit
$R=R_o$ is too high  for the magnetic field to be dynamically
important, compared with the gravitational interaction.

Now, we come back to the observational fact reported by
\citet{1997A&A...318..416P}, that the stellar rotation curve is markedly
lower in the outer disc region than the rotation curve of gas.
This is also true for the rotation curves in \figref{fig:RC},
which
{shows that the separation effect is not changed by
considering} an extended sample of stars {and by applying an
independent averaging method}. The South-North asymmetry in the
rotation of gas has appeared too. In addition, a South-North
asymmetry in the motion of stars is visible, but it is lower than
that for gas and has variable sign -- higher velocities closer to
$R_o$ are observed for stars on the Southern side, then, for
greater $R$, on the Northern side.

{The dispersion measure $S_V(R)$ of
observed azimuthal velocities of kinematical tracers about the
mean value $E_V(R)$, defined earlier and shown in \figref{eq:gest}
separately for stars and for gas, is comparable with the
separation of the stellar and gaseous rotation curves. It is true
that the separation may be entirely due to measurement
inaccuracies, and \citet{1997A&A...318..416P} discuss a possible
explanation, but the separation is also markedly large. The
dilemma can be solved only by increasing the accuracy of
 measurements. Our
hypothesis is that the whole effect (or its significant part) may
be caused by the presence of large scale magnetic fields.
 Then, important
is the real amount of the separation, although it might slightly
differ from that seen in \figref{eq:gest}. }

\begin{figure}[h]
   \centering
   \includegraphics[width=0.475\textwidth]{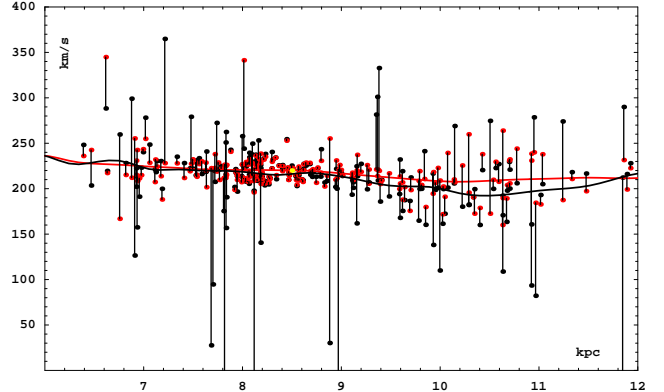}
      \caption{\label{fig:CephCompar} Rotation
curves for cepheids with measured radial and transversal velocity
components: {\it [red]--} true azimuthal component of
Galactocentric velocity and a moving average fit obtained based on
all three heliocentric components (the transformation parameters
were constrained so that the circular velocity is $220\,\km/\sec$
at $8.5\kpc$);  {\it [black]--} a re-projected azimuthal component
and a moving average fit obtained based only on the heliocentric
radial motion assuming the same transformation parameters as
before.   }
  \end{figure}

Before proceeding further, it should be tested, if the lack of
complete information about the transversal motions for a number of
heliocentric position-velocity data, may be consequential for the
shape of the obtained stellar rotation curve, and sufficient to
account for the observed difference in the rotation of stars and
gas. A fraction of cepheids in the sample used to prepare the
stellar rotation curve have all their three velocity components
measured.  We obtained for them two test rotation curves: one
based on all three heliocentric velocity components, and the other
based only on the radial heliocentric component (as if the
transverse heliocentric components were unknown). The result is
shown in \figref{fig:CephCompar}.
There is a difference seen for the test curves, but it is very
small in comparison to the {separation} between the stellar and
gaseous rotation curves.

\section{Gas density}

Magnetic field can not directly influence the motion of neutral
gas. The ionized gas fraction of the interstellar medium is more
diluted than the non-ionized fraction. But, as explained in the
introduction, it is dense enough for the magnetic field to be
frozen-in and to hold together the mixture of gases (including
molecular gas). The gas then moves as a whole driven by magnetic
tension. Its column mass density $\sigma(R)$ can be thus reliably
approximated by that of the neutral hydrogen. The latter we adopt
from \citep{2016PASJ...68....5N} and, re-project to a volume density of the
form:
\begin{equation}\label{eq:gest}
\varrho(R,{Z})=\frac{\varrho_{o}(R)}{\cosh^2\br{\frac{1}{2}(Z/h)}},
\qquad \varrho_o(R)=\frac{\sigma(R)}{4h}.\end{equation} For
$\abs{Z}\gg h$, the {profile function
$\cosh^{-2}\br{\frac{1}{2}(Z/h)}$} behaves as {$\exp\br{-|Z|/h}$},
however {$\varrho(R,Z)$} is smoother and half as high in the
{Galactic plane} vicinity as the cuspy exponential profile
{$\frac{\sigma(R)}{2h}\exp\br{-|Z|/h}$ with the same integrated
mass}.  The $\cosh^{-2}$ profile is a solution of the Jeans
equation, under the assumption of negligible horizontal variations
in the density, and almost uniform gravitational acceleration. The
disc thickness is related to  a width-scale parameter $h$, which
is {regarded as a single free parameter,} assumed to be
independent of $R$  (a region $-h<z<h$ comprises ${\sim}46\%$ of
total mass).

{The above volume density model with} constant thickness is {only
a coarse grained} approximation used to map a given column density
$\sigma(R)$ to a central volume density $\varrho_o(R)$. Real
{disc} thickness {grows quickly in an exponential fashion} with
the Galactocentric distance \citep{2007A&A...469..511K},
{moreover, the gas disc warps \citep{1997A&A...327..325M} with
asymmetric amplitude of a few $\kpc$
\citep{2006PASJ...58..847N,2006ApJ...643..881L}. Thus, it is not a
priori obvious how the single model parameter $h$ relates to the
real variable thickness of the warped disc. Moreover, our
simplified magneto-hydrodynamical model (which we discuss in more
detail later) neglects the vertical structure, thus the decrease
in the density with $\abs{Z}$ occurring when $h$ is very low must
be accounted for. The \hii regions which are used to determine the
Galactic rotation curve and whose motion is influenced by magnetic
fields, are located within a strip $|Z|<450\,\pc$ in which we
assume that the fields and volume density are independent of $Z$.
Now, for $h$ large enough, say  $h=1.5\kpc$ or more, the {volume}
density would change insignificantly within the strip and could be
assumed constant (equal to the central volume density), however,
for small $h$, say $h=450\pc$ or less, the decrease of the density
could not be neglected (the easiest way would be to take in place
of the central density some reduced, say, average density within
the strip).
 }

\begin{figure}[h]
   \centering
   \includegraphics[width=0.475\textwidth  ,trim={1cm 0cm 0cm 0cm}]{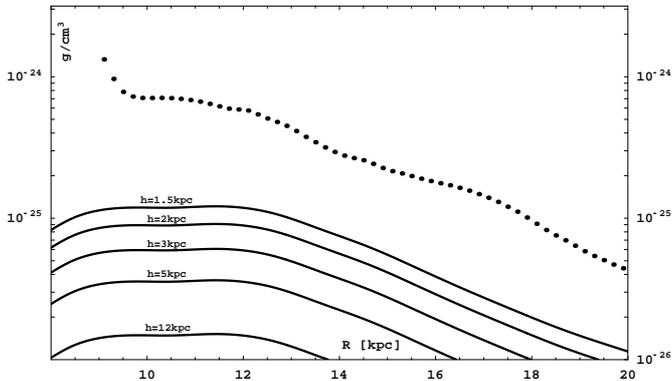}
      \caption{\label{fig:gestkapelusz}  {Central} volume density of the gaseous fraction \new{(}at $Z=0$\new{)}
corresponding to the density profile in \eqref{eq:gest}, shown for
various values of width-scale parameter $h$ {[{\it solid
lines}]. For comparison, the [{\it dotted line}] shows a volume
density at $Z=0$ from \citep{2009ARA&A..47...27K}.}}
  \end{figure}

 In \figref{fig:gestkapelusz}, the {central volume} density (at $Z=0$) {corresponding to \eqref{eq:gest},} is
shown for various $h$.
{In {the same figure}}
is shown also {a central} volume density from
\citep{2009ARA&A..47...27K}. As is seen, it is much higher than
considered by us. We stress, that surface density in
\citep{2016PASJ...68....5N} (based on which we obtain our volume
densities) and surface density in \citep{2009ARA&A..47...27K} are
very similar, therefore the difference in the volume densities
results from differences in the assumptions made about the {shape
and thickness of the} vertical profile of gas.
\citet{2009ARA&A..47...27K} predict {much} lower { half width at
half maximum scale height} than ours: in a region between $R_o$
and $2R_o$, {it} grows from $\approx 150\pc$ to $\approx 360\pc$,
see Fig.6 in \citet{2009ARA&A..47...27K}.

\section{Magnetic fields}

A magnetic field, necessary to account for the measured difference
in the circular velocities, can be obtained by finding a solution
of the stationary Navier-Stokes equation for an inviscid and
pressureless medium:
\begin{equation}\label{eq:navier}
\br{\vec{v}\circ\vec{\nabla}}\vec{v}=-\vec{\nabla}U +
\frac{1}{4\pi \varrho}\br{\vec{\nabla}\times\vec{B}}\times
\vec{B}.
\end{equation}
The field $\vec{B}$ can be decomposed into a series of modes,
lower modes for larger scale structures and higher modes for lower
scale structures; each mode with its own amplitude, possibly  some
of them dominating all the others. The dominant part of the
large-scale magnetic field of the Galaxy is likely to be
axisymmetric, according to the turbulent dynamo theory and in
agreement with observations \citep{1991ApJ...366..450V}. Under
axial symmetry and in the {Galactic} plane vicinity (where the
vertical component $B_{Z}$ and its first derivative can be assumed
to vanish by reflection symmetry with respect to that plane), the
system of equations \eqref{eq:navier} reduces to
\begin{subequations}\label{eq:navier12}
\begin{eqnarray}
\label{eq:navier1} &&V_{R}\,\frac{\partial V_R}{\partial R}=
\frac{\delta\, v_{\Phi}^2}{R}-\frac{1}{4\pi \varrho}\,
\frac{B_{\Phi}^2}{R} -\frac{1}{4\pi \varrho}\,
B_{\Phi}\frac{\partial B_{\Phi}}{\partial R},\\
 \label{eq:navier2}
&&V_{R}\frac{
\partial\br{{R}\,V_{\Phi|G}}}{\partial R}= \frac{1}{4 \pi
\varrho}\,B_{R}\frac{
\partial\br{{R}\,B_{\Phi}}}{\partial R}.
\end{eqnarray}
\end{subequations}
Here, $\delta\, v_{\Phi}^2=V_{\Phi |
G}^2-V_{\Phi | S}^2$ stands for a given difference of squares of
circular velocities $V_{\Phi | G}$ and $V_{\Phi | S}$,  that of
gas and stars, respectively, which is assumed to be of
non-gravitational origin and which, for the time being, we
attribute to the interaction with magnetic field. We will see to
what magnetic field magnitudes this hypothesis would lead to.

{As something of an aside, let's get some idea
about the role of Alfv{\'e}n speed
$V_A=\sqrt{\vec{B}\circ\vec{B}/4\pi\varrho}$ in our context. To
simplify things, it is worth considering purely circular orbits
when $\partial_{R}B_{\Phi}=0$, in  which case \eqref{eq:navier1}
implies $$\delta\, v_{\Phi}^2=\frac{B_{\Phi}^2}{4\pi \varrho}.$$
Then, the correction to $V_{\Phi | G}^2$ would be numerically
equal to the Alfv{\'e}n speed squared (if $B_{R}=0$) or to a
fraction of it (if $B(R)\neq0$). In a more interesting case with
circular orbits, when $\partial_{R}B_{\Phi}\neq0$, there is
additional term proportional to the gradient $\partial_RB_{\Phi}$
in \eqref{eq:navier1} that can increase (if
$\partial_{R}B_{\Phi}>0$) or reduce (if $\partial_{R}B_{\Phi}<0$)
the contribution from Alfv{\'e}n term to $\delta\, v_{\Phi}^2$, or
even change the sign of $\delta\, v_{\Phi}^2$, reducing the
circular velocity to values lower than the $V_{\Phi | G}$ value
valid in absence of magnetic field. For noncircular orbits, when
$V_R\neq0$, the situation becomes more complicated.}

The idealizations above, do not strictly reflect the reality and
have some drawbacks. By the assumed symmetries, $B_{Z}(0)=0$ and
$\partial_Z B_{Z}(0)=0$, and so $B_{Z}$ could be neglected  in the
{Galactic} plane vicinity. Then, the law
$\vec{\nabla}\circ\vec{B}=0$ would imply, for an axi-symmetric
field, a superposition of: a purely azimuthal field being an
arbitrary function of $R$, and a radial field of the form
$B_{R}=\mathrm{const}\cdot R^{-1}$ -- a limitation which seems not
much realistic. However, we may allow for a small violation of the
law $\vec{\nabla}\circ\vec{B}=0$, because $\vec{B}$ in
\eqrefs{eq:navier12} does not strictly describe the total magnetic
field.  Furthermore, axial symmetry, although very convenient,
holds only approximately;  the large-scale field will neither be
purely axisymmetric, nor reflection-symmetric, and the turbulent
part will be devoid of all the symmetries, etc. But there will
always be some correction to the simplified field, that restores
the divergence-less of the exact total field. The purpose of our
model is only to approximately determine the leading structure of
the total horizontal  magnetic field, and estimate its order of
magnitude, consistently with the observed difference of rotation,
without looking into the more sophisticated problem of modelling
this structure in detail. {To sum up, the assumption of axial
symmetry and of vanishing $B_Z$, is a simplification that allows
to solve equations. We neglect the limitation $B_R\sim R^{-1}$
implied by the constraint $\vec{\nabla}\circ\vec{B}=0$ -- although
consistent with the earlier simplification, the condition would be
too stiff. We consider it more realistic an approximate field
which is not quite divergence-free, remembering that the violation
is a result of necessary simplifying assumptions {(the
contribution to $\vec{\nabla}\circ\vec{B}$ introduced by $B_R$
could be balanced by a higher order correction or, even more
simply, by an antisymmetric $B_Z$ which is zero at the {symmetry}
plane and, therefore, not contributing to the radial Lorentz
force). }

For the reasons described above, we may assume a simple structure
of $\vec{B}$ in the disc plane vicinity admissible by axial and
reflection symmetry:
\begin{equation}
\label{eq:ansatz}
\{B_{R} ,B_{\Phi}, B_{Z}\}=B(R)\{\sin{\varepsilon},\eta
\cos{\varepsilon}, 0\}, \quad\eta=\pm1,\, \abs{\varepsilon}<\frac{\pi}{2}.
\end{equation}
In this ansatz $B(R)$ is assumed to be a positive function, then
various directions of $\vec{B}$ are realized by means of
parameters $\varepsilon$ and $\eta$. Here, $\varepsilon$ may be
considered as a small perturbation parameter. The law
$\vec{\nabla}\circ\vec{B}=0$ is violated by a term
$\sin{\varepsilon}R^{-1}\partial_R(R B(R))$, which can be
neglected when $\varepsilon$ is small enough or/and when $B(r)$ is
close to a function $\mathrm{const}\cdot R^{-1}$.

The form of \eqref{eq:navier} involves products of components of
$\vec{B}$. Hence, the residual rotation $\delta\, v_{\Phi}^2$ is
insensitive to any change in the direction of $\vec{B}$;
similarly, the reduced equations \eqrefs{eq:navier12} show that
$\delta\, v_{\Phi}^2$ will be preserved when the sign of
$B_{\Phi}$ is altered. For $\varepsilon=0$ there is no radial
component: $B_r=0$ and $V_R=0$, the solutions are purely
azimuthal. This mimics the leading large-scale azimuthal magnetic
field accounting for the $\delta\, v_{\Phi}^2$. For greater
$\varepsilon$, but still small, we obtain a perturbation of the
previous solution -- except for the leading azimuthal field
$B_{\Phi}$, a small radial field $B_R$ appears, which gives rise
to a perturbation of the radial velocity component. This mimics a
radial velocity dispersion one may expect to appear in the
presence of turbulent $B_R$: when the sign of \corr{$\varepsilon$}
is altered (this is equivalent to a composition of two
reflections: $\vec{B}\to-\vec{B}$ followed by
$B_{\Phi}\to-B_{\Phi}$), then $V_R$ also changes sign, while the
leading component $B_{\Phi}$ is still the same.   Note also that
the magnitudes of turbulent fields in the Galaxy may be comparable
or even exceeding the magnitudes of large-scale fields. The ansatz
\eqref{eq:ansatz} may be thus regarded as taking into account both
the large-scale purely azimuthal field with variable sign, as well
as a turbulent field - independent perturbations to both
components $B_{\Phi}$ and $B_{R}$.

Above, we have sketched our motivation behind the ansatz
\eqref{eq:ansatz} and its possible interpretation. We also draw to
attention the fact that \corr{the ansatz} is a special case of a
field configuration $B(R)\{\sin(p)
\cos{\chi(Z)},\cos(p)\cos{\chi(Z)},\sin{\chi(Z)}\}$, considered in
the context of describing Galactic magnetic fields
\citep{2010A&A...522A..73R} (with $\sin{p}\ne0$ the latter would
be divergence free  for $B(R)=B_o\frac{R_o}{R}\exp{\br{-\frac{k\,
R}{\sin{p}}}}$ and $\chi(Z)=k Z+\chi_o$). In particular, the
result in \citep{bib:polema2} shows that \eqref{eq:ansatz}
describes well an axisymmetric spiral pattern of $3\muG$, observed
in the regular disc field of Galaxy between $3$ and $20\kpc$ (then
$p\approx15^{o}$).

\medskip\noindent
{\it Solutions.} {In the region where $\partial_R
\br{RV_{\Phi|G}}\neq0$,} the second of equations in
\eqrefs{eq:navier12} can be solved for $V_R$ and then substituted
in the first equation. This gives a nonlinear second order
ordinary differential equation for $B(R)$ with coefficients being
functions of known functions $\varrho(R)$, $V_{\Phi|G}$ and
$V_{\Phi|S}$, and their derivatives. We solved this equation using
numerical integration. After \citet{bib:polema2}, we put
$\varepsilon\approx\frac{\pi}{12}$. The resulting magnitude of
magnetic field, $B(R)$, accounting for the observed difference
between the rotation of gas and stars, is shown versus the
Galactocentric distance in the {Galactic} plane in
\figref{fig:polemag1}, for three values of parameter $h$.
\begin{figure}[h]
   \centering
   \includegraphics[width=0.7\textwidth,clip,trim={1.5cm 5cm 0.4cm 1.5cm}]{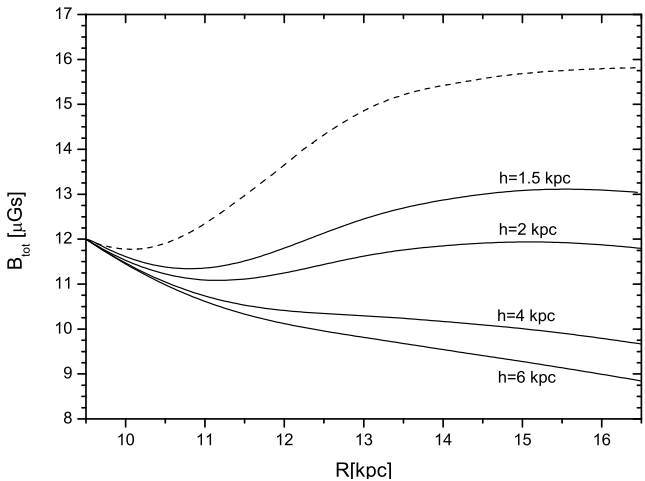}
      \caption{\label{fig:polemag1}
The modelled magnitude of magnetic field that accounts for the
observed difference between the stellar rotation curve and the
rotation curve of partially ionized gaseous medium. The {four
solid} curves correspond to various values of the width-scale
parameter $h$ of the model gaseous disc.  {The {\it diamond line}
shows magnetic field corresponding to a volume density of
\citep{2009ARA&A..47...27K} corrected for the altitudes of \hii
regions.} }
  \end{figure}
It follows, that $B(R)$ depends on the disc thickness, and hence
{is} related to the density of gas in the {disc} plane. For the
considered range of $h$ values between {$1.5\kpc$} and $6\kpc$,
the required $B(R)$ fall within a range between $8.5\muG$  and
{$13\muG$}. {For a volume density as published in
\citep{2009ARA&A..47...27K}, the corresponding $B(R)$ is higher,
between $\approx11.8\muG$ and $\approx15.8\muG$.} For comparison,
in \figref{fig:pole2} similar results are shown, accounting for
the difference between the rotation of stars and gas, obtained
when the rotation curve of gas is split into two parts: one for
the gas above the {Galactic} plane, and the other for the gas
below {that} plane.
\begin{figure}[h]
   \centering
      \includegraphics[width=0.7\textwidth,clip,trim={1.5cm 5cm 0.4cm 1.5cm}]{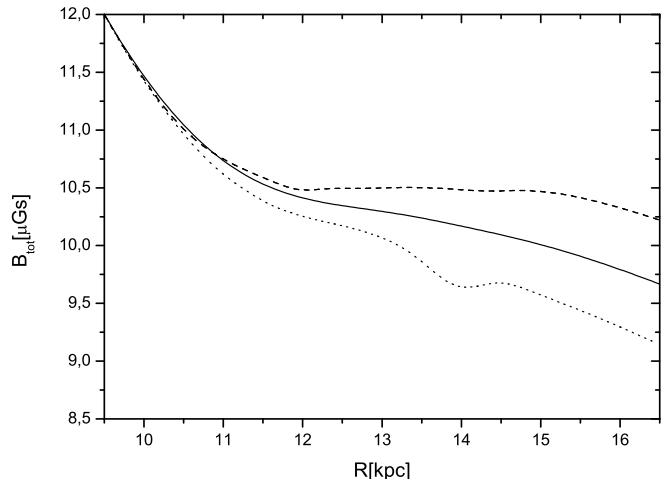}
      \caption{\label{fig:pole2} Comparison of the modelled magnetic
field magnitudes that would account for the difference between the
stellar rotation curve and three gaseous rotation curves obtained
in cases when: 1) only the gas below the Galactic plane is
considered ({\it dashed line}), 2) only the gas above the Galactic
plane is considered ({\it dotted line}) and 3) when whole of the
gas, that is,  both below and above that plane is taken into
account. The results were obtained assuming $h=4\kpc$. }
  \end{figure}

{In \figref{fig:alfven} we compare (in units of speed) the contributions of magnetic
terms present in the azimuthal part of equations
\eqref{eq:navier12}. The square root of Alfv{\'e}n term
$B_{\Phi}^2/4\pi\varrho$ (comparable with the Alfv{\'e}n speed
$\sqrt{\vec{B}\circ\vec{B}/4\pi\varrho}$ for the particular
solution), grows with radius, reaching values comparable with the
circular velocity of gas. The square root of absolute value of the
gradient term $B_{\Phi}\partial_R
      B_{\Phi}/4\pi\varrho$, reaches values roughly twice lower, and depending on
the sign, it increases or decreases the effect from the Alfv{\'e}n
term.}
\begin{figure}[h]
   \centering
      \includegraphics[width=0.7\textwidth,clip,trim={1.5cm 5cm 0.4cm 1.5cm}]{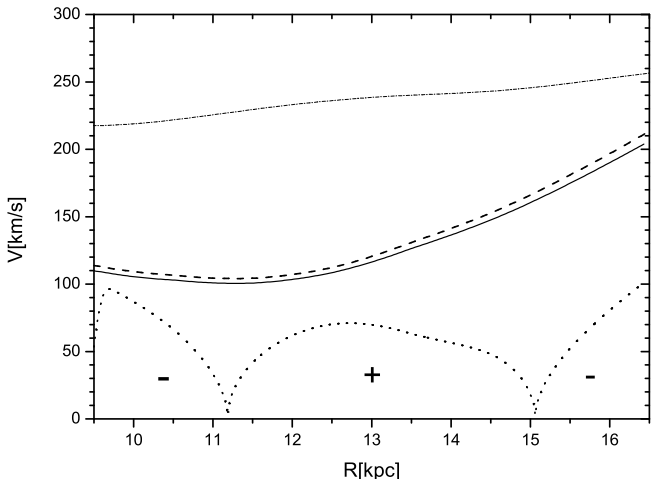}
      \caption{\label{fig:alfven} {Comparison of magnetic terms in equation
      \eqref{eq:navier1} for a solution with $h=2\kpc$. Square root of Alfv{\'e}n term
      $B_{\Phi}^2/4\pi\varrho$ [{\it solid line}], square root of
      absolute value of a gradient magnetic term $B_{\Phi}\partial_R
      B_{\Phi}/4\pi\varrho$ [{\it dot line}]
      (regions where this term is positive or negative are marked $+$ or
      $-$). The Alfv{\'e}n speed
      $\sqrt{\vec{B}\circ\vec{B}/4\pi\varrho}$ [{\it dashed line}]
      is comparable in magnitude with the circular speed of gas [{\it dash-dot
      line}].}}
  \end{figure}

{The radial velocity component $V_R$ can be obtained for the found
solutions from \eqref{eq:navier2}. In \figref{fig:VR} the velocity
is shown for a particular solution with $h=2\,\kpc$, however the
velocity is almost the same for other $h$, the difference is
noticeable only in the boundary region close to $R=9\kpc$.  }
\begin{figure}[h]
   \centering
      \includegraphics[width=0.7\textwidth,clip,trim={1.5cm 5cm 0.4cm 1.5cm}]{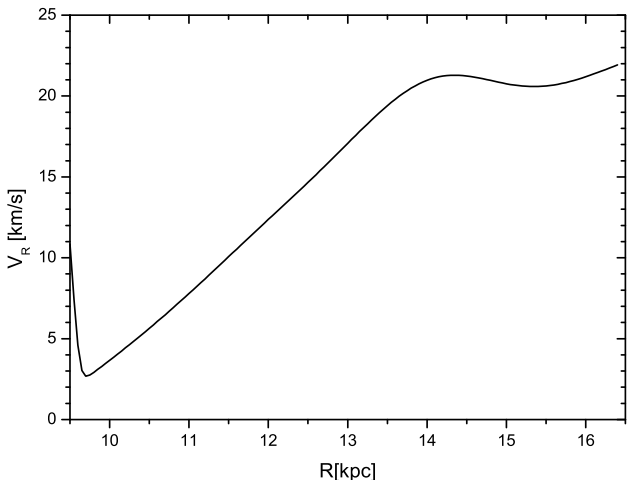}
      \caption{\label{fig:VR} {Radial velocity component in Galactic magnetic field, estimated based on equation \eqref{eq:navier2} assuming $h=2\,\kpc$.} }
  \end{figure}
{As we have seen earlier, the $V_R$ component can be interpreted
in our model as a measure of the radial velocity dispersion due to
the interaction with magnetic field that flops its radial
direction, in which case the sign $V_R$ changes following these
flops. The $V_R$ is of similar order of magnitude as the radial
velocity dispersion observed for spiral galaxies, see
\cite{2009AJ....137.4424T}, what suggests that magnetic fields may
be important in modeling the dispersion of gas. One can also use
the $V_R$ component to estimate the accretion rate connected with
the radial flow of gas. The accretion rate is defined in our case
as the amount of matter flowing through a cylindric surface of
radius $R$. Accordingly, using the density profile
\eqref{eq:gest}, it can be estimated from $2\pi
R\,\Sigma(R)V_R(R)$ or $8\pi R\,h\varrho(R,0)V_R(R)$.  The
resulting accretion rate for the same solution as above is shown
in \figref{fig:AccRate}, it will similarly be almost independent
of $h$. }
\begin{figure}[h]
   \centering
      \includegraphics[width=0.7\textwidth,clip,trim={1.5cm 5cm 0.4cm 1.5cm}]{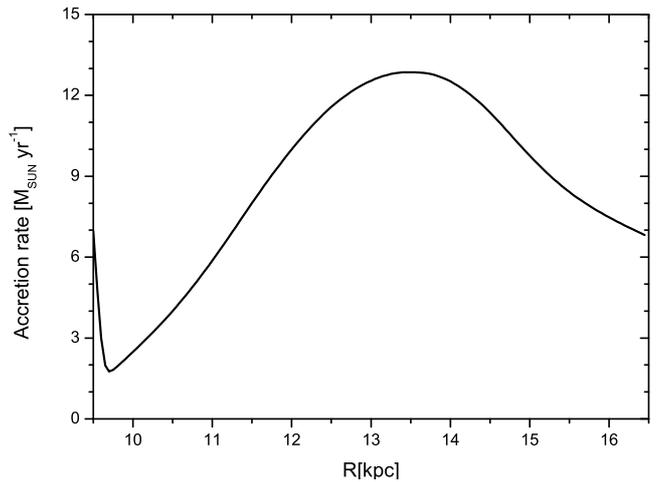}
      \caption{\label{fig:AccRate} {Accretion rate associated with the radial velocity component shown in \figref{fig:VR}.} }
  \end{figure}

\section{Conclusions}
The {difference} between the stellar and gaseous rotation curves
observed in the outer Galactic disc \citep{1997A&A...318..416P} is
significant. Attributing it entirely to a magnetic interaction
would require fields exceeding $10\muG$, strongly depending on the
gas density in the Galactic disc. The question arises, whether
fields of this magnitude could be present in the Galaxy. According
to the interpretation sketched before, the ansatz
\eqref{eq:ansatz} for the magnetic field we assumed for solutions
of the Navier-Stokes equations, takes into account the influence
of the total magnetic field, including both the regular and
turbulent magnetic components.  It is important to note that the
magnitude of the turbulent component in the Galaxy is not
negligible, and may exceed the magnitude of the regular component.
The regular field observed in the Galaxy is estimated to have a
value of $5\muG$, while the turbulent field has $11\muG$
\citep{bib:magp2}. Furthermore, \citet{bib:magp2} estimate, based
on their models, that each of the components $B_{R}$, $B_{\Phi}$
and $B_{Z}$ of the total field may reach local values as high as
$15\muG$  or more for lower radii, falling off with the radial
distance, and exceeding $10\muG$, even at the radius of the solar
orbit. Our results seem consistent with these findings.

{ For lower densities (higher scale parameters $h$), the magnetic
field, as seen in \figref{fig:polemag1}, is a decreasing function
of radius, while for higher densities (corresponding to
$h=2\,\kpc$, $1.5\,\kpc$ or lower) the field increases with radius
or is approximately constant. Most often, the magnetic field is
expected to decrease exponentially \citep{bib:magp2}, which may
seem to disagree with our results at higher densities. Howewer, it
should be remembered that the real density of \hii regions, used
to determine the gas rotation curve, may be lower than the density
obtained for \hi distribution. This is because ionized gas, such
as in \hii regions, should have lower density than the neutral
hydrogen \hi. If this was the case, the densities corresponding to
higher $h$ (above $2\,\kpc$) would better approximate the density
of the \hii regions, which are influenced by magnetic fields.
Then, the magnetic field found by us would have values and
behaviour consistent with the usual assumptions. }

{Interestingly, we observed that Alfv{\'e}n velocity reaches
values comparable with the circular velocity of gas. Alfven
velocities of more than $100\,\km/\mathrm{s}$ are about 10 times
larger than the mid-plane turbulent velocity of
$10\,\km/\mathrm{s}$. It means a $100$ times higher magnetic
energy compared to the kinetic energy of the turbulence. }

{The volume gas density, important for the magnetic field
determination, is strongly dependent on the assumed vertical
profile of the gas layer. Fortunately, this uncertainty seems not
so essential to the modelled magnetic field magnitudes required by
the observed separation of stellar and gaseous rotation curves.
This is so, because a higher volume density in the Galactic plane
means a lower postulated scale height $h$, which in turn requires
taking into account the altitudes of the \hii regions used to
determine the rotation curve of gas. The ionized hydrogen of \hii
regions is much more diluted than the neutral hydrogen of \hi
regions. Both kinds of gas, \hi and \hii, may belong to the same
gas cloud which must move as a whole. It is therefore safer (in
order not to reduce the magnetic field too much) to determine the
(ionized) gas density such as if the gas consisted of \hi hydrogen
only. This approach overestimates the density of the ionized
fraction.}

As seen in \figref{fig:pole2}, the difference between circular
velocities of gas in the northern and southern sides of the
Galactic disc, if attributed to magnetic forces, requires small
difference in the magnetic field magnitudes, not greater than
$1.3\muG$. It is thus plausible that the difference in rotation
may indeed be due to some asymmetry in the distribution of
magnetic field.

\medskip
To sum up: as the above magnetic field estimates show, it is
plausible that most of the observed difference in the circular
velocities of the stellar and gaseous fractions in the Galaxy may
be caused by the presence of interstellar magnetic fields, frozen
in the partially ionized gas, and holding the gaseous mass
component together by the resulting magnetic tension. { Within the
possible range of higher densities corresponding to $h$ about
$1-2\,\kpc$ or less, the magnetic field needs to increase with the
distance (it is bound within $12$ and $16\,\muG$) or at least to
be approximately constant between $10$ and $20\,\kpc$. Only for
very low gas density at the Galactic plane, lower than
$10^{-25}\,\mathrm{g}/\mathrm{cm}^{3}$, the magnetic field is
decreasing with the Galactocentric distance, as observations
suggest; the difference in rotation could be then explained by
magnetic fields of order $10\,\muG$. If the real fields are
different, then only a fraction of the rotation difference could
be explained by the presence of magnetic fields.} {In any case,}
the difference in the rotation could be used as a means to
estimate the intensity of the total magnetic field in spiral
galaxies. This is also a manifestation of the influence of
magnetic fields on the dynamics of spiral galaxies. Apart form the
influence on the rotation, the radial component of magnetic field
will modify the radial velocity dispersion.

\section*{Acknowledgments}
We would like to thank the referee for carefully
reading the manuscript and for various constructive suggestions.

\nopagebreak
\bibliography{MilkyWayMagnetic_arxiv_v1}
\bibliographystyle{apalike}

\end{document}